\documentclass[prl,letterpaper,aps,superscriptaddress,floatfix,twocolumn]{revtex4-1}
\usepackage{graphicx}
\usepackage{amsmath}
\usepackage{amsfonts}
\usepackage{subfigure}
\newcommand{\beq}{\begin{equation}}
\newcommand{\eeq}{\end{equation}}
\newcommand{\bk}{{{\bf{k}}}}

\newcommand{\bB}{{\bf{B}}}
\newcommand{\bE}{{\bf{E}}}

\newcommand{\bs}{{\bf{s}}}
\newcommand{\bj}{{\bf j}}
\newcommand{\beqa}{\begin{eqnarray}}
\newcommand{\eeqa}{\end{eqnarray}}
\newcommand{\pdg}{{\vphantom \dag}}
\newcommand{\dg}{{\dag}}
\newcommand{\bnabla}{{\boldsymbol \nabla}} 
\newcommand{\bsigma}{{\boldsymbol \sigma}}

\newcommand{\upa}{\uparrow}
\newcommand{\da}{\downarrow}

\newcommand{\cL}{{\cal L}}
\begin{document}
\title{$\mathbb Z_2$ and Chiral Anomalies in Topological Dirac Semimetals}
\author{Anton A. Burkov}
\affiliation{Department of Physics and Astronomy, University of Waterloo, Waterloo, Ontario 
N2L 3G1, Canada} 
\affiliation{ITMO University, Saint Petersburg 197101, Russia}
\author{Yong Baek Kim}
\affiliation{Department of Physics and Center for Quantum Materials, University of Toronto, Toronto, Ontario M5S 1A7, Canada}
\affiliation{Canadian Institute for Advanced Research, Toronto, Ontario M5G 1Z8, Canada}
\date{\today}
\begin{abstract}
We demonstrate that topological Dirac semimetals, which possess two Dirac nodes, separated in momentum space along a rotation axis and protected by rotational symmetry, exhibit an additional quantum anomaly, distinct from the chiral anomaly. 
This anomaly, which we call the $ \mathbb Z_2$ anomaly, is a consequence of the fact that the Dirac nodes in topological 
Dirac semimetals carry a $\mathbb Z_2$ topological charge. The $\mathbb Z_2$ anomaly refers to nonconservation of this charge in the presence of external fields due to quantum effects and has observable consequences due to its interplay with the chiral anomaly. 
We discuss possible implications of this for the interpretation of magnetotransport experiments on topological Dirac 
semimetals. 
We also provide a possible explanation for the magnetic field dependent angular narrowing of the negative longitudinal 
magnetoresistance, observed in a recent experiment on Na$_3$Bi. 

\end{abstract}
\maketitle
The recent theoretical~\cite{Wan11,Ran11,Burkov11-1,Burkov11-2,Xu11,Krempa12,Kane12,Fang12,Fang13}
and experimental~\cite{Chen14,Neupane14,HasanTaAs,DingTaAs2,DingTaAs,Lu15}
discoveries of Weyl and Dirac semimetals have extended the growing family of materials with topologically nontrivial 
electronic structure. 
They have also highlighted beautiful analogies and connections that exist between the physics of materials with topologically nontrivial 
electronic structure and the physics of relativistic fermions, described by the Dirac equation, which were anticipated some time 
ago by Volovik and others~\cite{Volovik88,Volovik03,Volovik07,Murakami07}.

Chiral anomaly, which refers to nonconservation of the chiral charge in the presence of collinear external electric and magnetic fields, is a particularly striking and important example. 
First discovered theoretically by Adler~\cite{Adler69} and by Bell and Jackiw~\cite{Jackiw69}, it provided the explanation 
for the observed decay of a neutral pion into two photons, prohibited by the chiral charge conservation, or chiral symmetry. 
Very recently, a condensed matter manifestation of the chiral anomaly was observed in a Dirac semimetal 
Na$_3$Bi~\cite{Ong_anomaly}, and, possibly, also in Weyl semimetal TaAs~\cite{Hasan_anomaly,Fang_anomaly} and in ZrTe$_5$~\cite{Kharzeev14}, which is proposed to be a Dirac semimetal. 

However, despite analogies, condensed matter systems with topologically nontrivial electronic structure are certainly significantly richer than the relativistic Dirac equation. 
In particular, there in fact exist two distinct classes of Dirac semimetals~\cite{Nagaosa14,Furusaki15}.
One, in which the Dirac points occur at time reversal invariant momenta in the first Brillouin zone (BZ)~\cite{Kane12}, 
and the second, in which the Dirac points occur in pairs, separated in momentum space along a rotation 
axis~\cite{Fang12,Fang13}. 
The Dirac semimetals, that have currently been realized experimentally~\cite{Chen14,Neupane14}, are of the second kind.
It is now understood~\cite{Nagaosa14,Furusaki15,Miransky15,Kee15,Sato15} that the Dirac points in such semimetals possess a nontrivial 
$\mathbb Z_2$ topological invariant, which protects the nodes and leads to the appearance of Fermi arc surface states, which 
connect projections of the node locations on the surface BZ, much like in Weyl semimetals. 

A natural question to ask in this regard is whether the existence of such a $\mathbb Z_2$ topological charge manifests in any way in transport, as the chiral charge of Weyl nodes manifests in negative longitudinal magnetoresistance, attributed to the 
chiral anomaly~\cite{Spivak12,Burkov_lmr_prl,Burkov_lmr_prb,Ong_anomaly}. 
In this paper we answer this question in the affirmative. 
We demonstrate that the Dirac semimetals with two Dirac nodes, carrying the $\mathbb Z_2$ topological charge, such as 
Na$_3$Bi and Cd$_2$As$_3$, exhibit the corresponding $\mathbb Z_2$ quantum anomaly, in addition to the chiral anomaly. 
We further demonstrate that the interplay of the two anomalies leads to observable manifestations in magnetotransport experiments.
We also discuss possible relevance of our results to the recent magnetotransport measurements on 
Na$_3$Bi~\cite{Ong_anomaly}, which have provided the first strong experimental evidence for the chiral anomaly
in condensed matter. 
In particular, we give a possible explanation for the magnetic field dependent angular narrowing of the negative longitudinal 
magnetoresistance due to the chiral anomaly, observed 
in this experiment. 

While in this work we will specifically focus on the case of Na$_3$Bi, our results are equally applicable to 
Cd$_2$As$_3$. 
We start from the low energy Hamiltonian of Na$_3$Bi in momentum space, derived in~\cite{Fang12}
\beq
\label{eq:1}
H = v_F (\sigma^x s^z k_x - \sigma^y k_y) + m(\bk) \sigma^z + \frac{\gamma}{2} \sigma^x  k_z (s^+ k_-^2 + s^- k_+^2). 
\eeq
The Pauli matrices $\bs$ and $\bsigma$ act on the spin and the orbital parity degrees of freedom correspondingly
and we will be using $\hbar = c = 1$ units throughout. 
The first two terms in Eq.~\eqref{eq:1} describe coupling between states of opposite parity, which forces them to be linear 
in the crystal momentum, measured from the $\Gamma$-point in the first BZ. 
Since $\sigma^z$ is the parity operator, the ``mass" term $m(\bk)$ is parity-even and has the following low energy form 
\beq
\label{eq:2}
m(\bk) = - m_0 + m_1 k_z^2, 
\eeq
which gives a pair of Dirac points at $\bk_{\pm} = (0, 0, \pm \sqrt{m_0/ m_1} \equiv \pm k_0)$. 
The last term in Eq.~\eqref{eq:1} is third order in the crystal momentum as a consequence 
of three-fold rotational symmetry of the crystal structure of Na$_3$Bi, where the rotation axis is the $z$-axis in Eq.~\eqref{eq:1}. 
This term is much smaller than the other  terms in the Hamiltonian in the vicinity of the Dirac points and we will thus ignore it henceforth. 
This omission has at most a quantitative effect on our results, but makes the presentation more transparent. 

We now make the following important observation. 
In the absence of the last term in Eq.~\eqref{eq:1}, the $z$-component of the spin is a strictly conserved quantity. 
With the last term included, it will no longer be strictly conserved, but will have a long relaxation time due to the smallness 
of this term, which is what ultimately justifies ignoring it physically. 
Let $s = \pm 1$ be the eigenvalues of $s^z$. 
Then the Hamiltonian separates into two independent $2 \times 2$ blocks, each describing a Weyl semimetal with a single 
pair of Weyl nodes, separated along the $z$-axis in momentum space
\beq
\label{eq:3}
H_s = v_F (\sigma^x s k_x - \sigma^y k_y) + m(\bk) \sigma^z. 
\eeq
The two Weyl semimetals are related to each other by the time reversal operation and thus each of the two 
Dirac band touching points at $\bk_\pm$ contains two Weyl nodes of opposite chirality and opposite eigenvalue $s$. 
It is convenient to expand the Weyl Hamiltonians $H_s$ near the nodes.
To linear order, one obtains
\beq
\label{eq:4}
H_s = v_F s \sigma^x k_x - v_F \sigma^y k_y + \tilde v_F \tau^z \sigma^z (k_z - \tau^z k_0), 
\eeq
where the two eigenvalues of the Pauli matrix $\tau^z$, $\tau = \pm 1$ refer to the two nodes and $\tilde v_F = 2 \sqrt{m_0 m_1}$. 
We now introduce Hermitian $4 \times 4$ gamma matrices as 
\beq
\label{eq:5}
\Gamma_s^1 \equiv \gamma^0_s \gamma^1_s = s \sigma^x, \,\, 
\Gamma_s^2 \equiv \gamma^0_s \gamma^2_s = - \sigma^y, \,\, 
\Gamma_s^3 \equiv \gamma^0_s \gamma^3_s = \tau^z \sigma^z, 
\eeq
where $\gamma^\mu_s$ are the relativistic Dirac gamma matrices. 
Note that we do not need an explicit representation for the Dirac matrices 
$\gamma^\mu_s$, all we really need to know are the Hermitian gamma matrices in Eq.~\eqref{eq:5}, the 
Dirac matrices are introduced only as a convenient notation. 
We may now define the chiral charge operator
\beq
\label{eq:6}
\gamma^5_s = i \gamma^0_s \gamma^1_s \gamma^2_s \gamma^3_s = - i \Gamma^1_s \Gamma^2_s \Gamma^3_s = - s \tau^z, 
\eeq
and the $\mathbb Z_2$ charge operator 
\beq
\label{eq:7}
s \gamma_s^5 = - \tau^z.
\eeq 
The physical meaning of the chiral charge operator is clear: the eigenvalues of $\gamma^5_s$ are the chiralities of the 4 Weyl fermions, which make up the two Dirac fermions. 
To clarify the meaning of the $\mathbb Z_2$ charge operator, we note that Eq.~\eqref{eq:7} is equivalent to 
$C_{\mathbb Z_2} = (C_\upa - C_\da)/2$, 
where $C_{\mathbb Z_2}$ is the $\mathbb Z_2$ charge and $C_{\upa, \da}$ are the chiral charges of the spin-up and
spin-down Weyl fermions. 
This definition is closely analogous to the definition of the $\mathbb Z_2$ invariant for a two-dimensional quantum spin Hall insulator with conserved spin in terms of the spin Chern numbers~\cite{Hasan10}.
Once the spin-conservation-violating terms in Eq.~\eqref{eq:1} and impurity scattering are included, only the modulo 2 part of the 
$\mathbb Z_2$ charge retains its meaning.
Note that both the chiral charge and the $\mathbb Z_2$ charge operators commute with the linearized Hamiltonian, expressing the 
approximate chiral and $\mathbb Z_2$ charge conservation.

\begin{figure}[t]
\subfigure{
\label{fig:1a}
  \includegraphics[width=8cm]{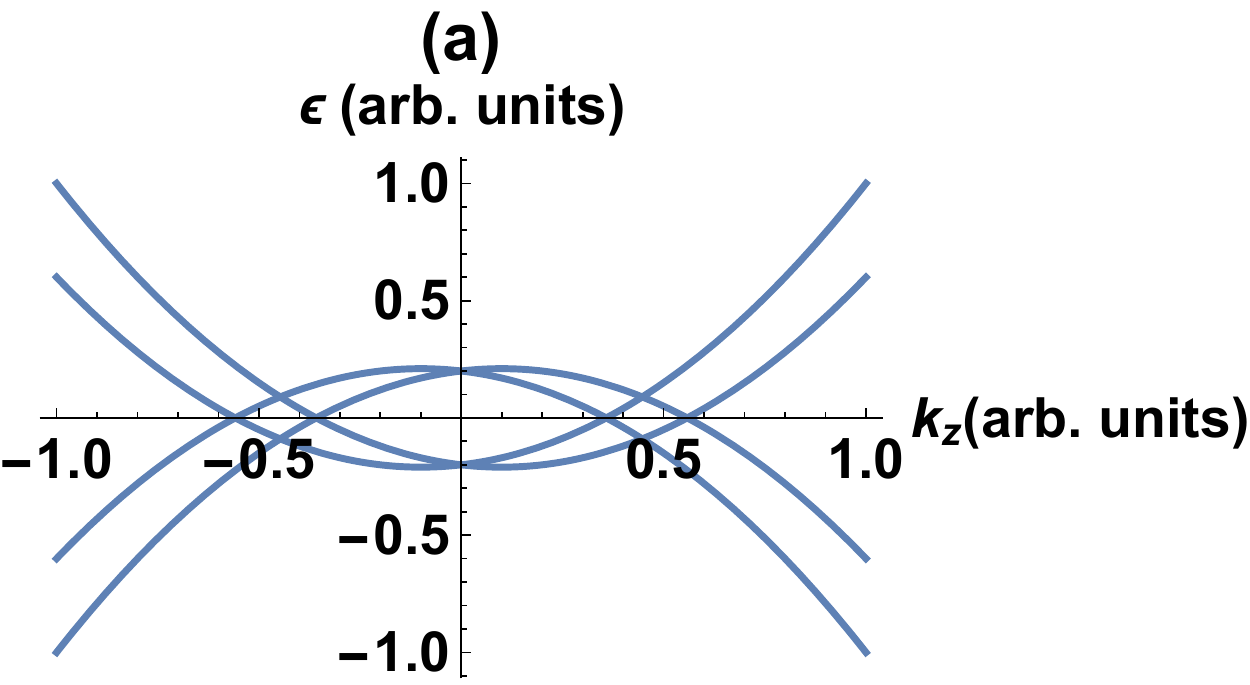}}
  \subfigure{
  \label{fig:1b}
  \includegraphics[width=8cm]{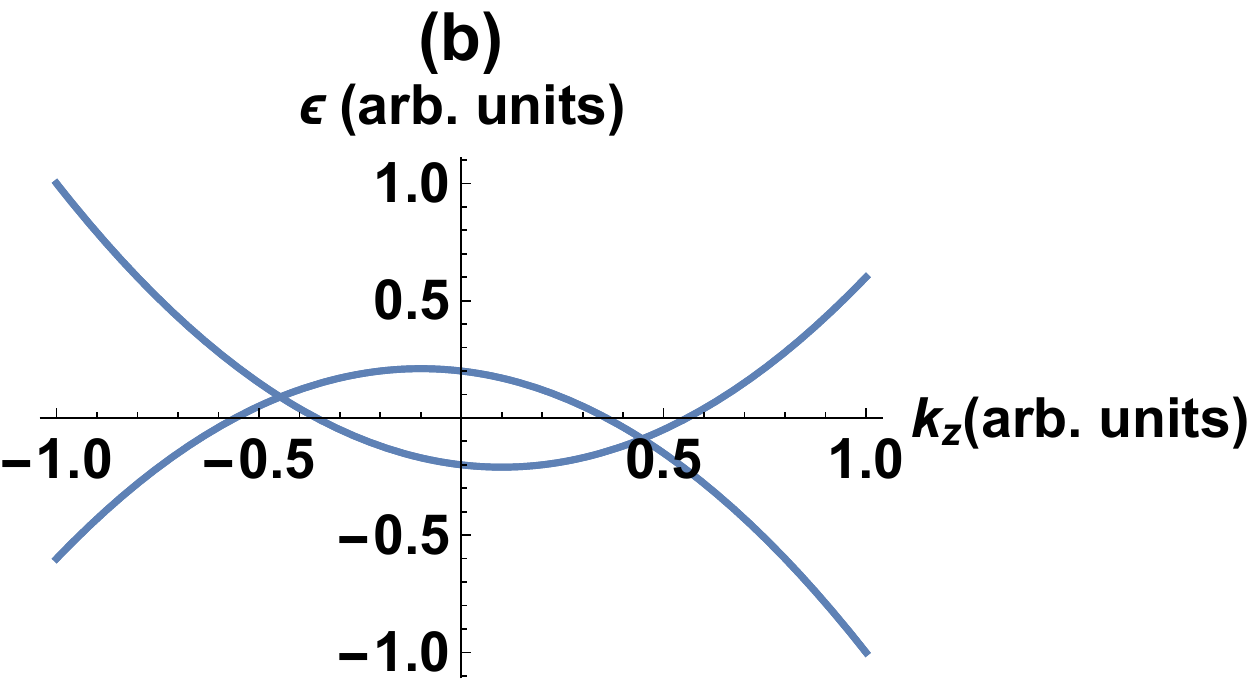}}
  \caption{(Color online) (a) Effect of a nonzero chiral chemical potential $\mu_5$ on the double Dirac electronic structure. 
  (b) Effect of a nonzero $\mathbb Z_2$ chemical potential $\tilde \mu_5$ on the electronic structure.}  
  \label{fig:1}
\end{figure}

What is particularly interesting is that the existence of the conserved $\mathbb Z_2$ charge has experimentally observable manifestations, which may be regarded as manifestations of the $\mathbb Z_2$ anomaly in analogy to the chiral anomaly. 
To see this, let us couple gauge fields to the fermions. 
Since we have two conserved quantities: charge and the $z$-component of the spin, we may introduce, for now purely formally, 
both the charge and the spin gauge fields. The real time Lagrangian, written in the relativistic notation, has the form
\beqa
\label{eq:8}
\cL&=&\psi_s^\dg i \partial_t \psi_s^\pdg - H \nonumber \\
&=&\bar \psi_s i \gamma^\mu_s [\partial_\mu + i e (A_\mu + s \tilde A_\mu) + 
i \gamma^5_s (b_\mu + s \tilde b_\mu)] \psi^\pdg_s, 
\eeqa
where summation over $s$ is implicit, $\bar \psi_s = \psi^\dg_s \gamma^0_s$, and we have absorbed the Fermi velocities into the definition of the corresponding coordinates. $A_\mu$ in Eq.~\eqref{eq:8} are electromagnetic gauge fields, which couple symmetrically to the fermions with different spin eigenvalue
$s$, while $\tilde A_\mu$ are the fictitious spin gauge fields, which couple antisymmetrically. 
What gives them physical meaning is that the functional derivative of the action with respect to $\tilde A_\mu$ produces the corresponding component of the spin current, which is well defined since the spin is conserved. 
$b_\mu$ are the chiral gauge fields, which couple antisymmetrically to Weyl fermions of opposite chirality and thus shift them 
in opposite directions in momentum space or in energy. 
Finally, $\tilde b_\mu$ couple antisymmetrically to fermions with different $\mathbb Z_2$ charge and thus may be called, with a slight 
abuse of terminology, the $\mathbb Z_2$ gauge fields. 
Specifically, we have
\beq
\label{eq:9}
b_\mu = (\mu_5, 0, 0, 0),\,\, \tilde b_\mu = (\tilde \mu_5, 0, 0, b), 
\eeq
where $b \equiv \tilde v_F k_0$. 
Here $\mu_5$ is the chiral chemical potential, which is conjugate to the chiral charge operator and shifts 
Weyl fermions of opposite chirality in opposite direction in energy. It is equal to zero in equilibrium, but 
will in general be nonzero away from equilbrium in the presence of charge currents and external electromagnetic 
field, see Fig.~\ref{fig:1}. 
Similarly, $\tilde \mu_5$ is the $\mathbb Z_2$ chemical potential, conjugate to the $\mathbb Z_2$ charge operator. 
It shifts the two Dirac points in opposite directions in energy. 
Finally, the $z$-component of $\tilde b_\mu$ is the only component that is present in equilibrium and simply determines 
the distance between the two Dirac points in momentum space.   

Integrating out fermions in Eq.~\eqref{eq:8} using e.g. the Fujikawa's method~\cite{Fujikawa79}, we obtain~\cite{Zyuzin12-1,Hughes15}
\beqa
\label{eq:10}
S&=&-\frac{e^2}{2 \pi^2}  \int dt d^3 r \,\, b_{\mu} \epsilon^{\mu \nu \alpha \beta} (A_\nu \partial_\alpha A_\beta + 
\tilde A_\nu \partial_\alpha \tilde A_\beta) \nonumber \\
&-&\frac{e^2}{2 \pi^2}  \int dt d^3 r \,\, \tilde b_{\mu} \epsilon^{\mu \nu \alpha \beta} (A_\nu \partial_\alpha \tilde A_\beta + 
\tilde A_\nu \partial_\alpha A_\beta). 
\eeqa
As mentioned above, functional derivative of $S$ with respect to the gauge field $A_{\mu}$ gives the charge current, while functional derivative with respect to $\tilde A_{\mu}$ gives the spin current
\beqa
\label{eq:11}
j^\nu&=&-\frac{\delta S}{\delta A_{\nu}} = \frac{e^2}{\pi^2} b_{\mu} \epsilon^{\mu \nu \alpha \beta}\partial_\alpha A_\beta
+ \frac{e^2}{\pi^2} \tilde b_{\mu} \epsilon^{\mu \nu \alpha \beta}\partial_\alpha \tilde A_\beta, \nonumber \\
\tilde j^\nu&=&-\frac{\delta S}{\delta \tilde A_{\nu}} = \frac{e^2}{\pi^2} b_{\mu} \epsilon^{\mu \nu \alpha \beta}\partial_\alpha \tilde A_\beta
+ \frac{e^2}{\pi^2} \tilde b_{\mu} \epsilon^{\mu \nu \alpha \beta}\partial_\alpha A_\beta. \nonumber \\
\eeqa
On the other hand, since $b_\mu$ and $\tilde b_\mu$ act as chiral and $\mathbb Z_2$ gauge fields correspondingly, 
functional derivatives of the action
with respect to these give the chiral and the $\mathbb Z_2$ currents. 
Upon taking the divergence, we then obtain the following anomalous chiral and $\mathbb Z_2$ charge conservation laws
\beqa
\label{eq:12}
\partial_\mu j^\mu_5 = \frac{e^2}{8 \pi^2} \epsilon^{\mu \nu \alpha \beta} (F_{\mu \nu} F_{\alpha \beta} + 
\tilde F_{\mu \nu} \tilde F_{\alpha \beta}), \nonumber \\
\partial_\mu \tilde j^\mu_5 = \frac{e^2}{8 \pi^2} \epsilon^{\mu \nu \alpha \beta} (F_{\mu \nu} \tilde F_{\alpha \beta} + 
\tilde F_{\mu \nu} F_{\alpha \beta}), 
\eeqa
where $F_{\mu \nu} = \partial_\mu A_\nu - \partial_\nu A_\mu$ and $\tilde F_{\mu \nu} = \partial_\mu \tilde A_\nu - \partial_\nu 
\tilde A_\mu$ are the field strengths. 
We may pick a gauge for the spin gauge fields $\tilde A_\mu$ in which only the temporal 
component $\tilde A_0$, which is conjugate to the spin density, is nonzero, as we do not expect spin analogs 
of magnetic fields to arise in our context. 
Henceforth we will thus take $\tilde A_\mu = (\tilde A_0, 0, 0, 0)$. 
The physical meaning of $\tilde A_0$ is defined by the relation $\tilde j^0 = g(\epsilon_F) \tilde A_0$, where 
$\tilde j_0$ is the spin density and $g(\epsilon_F)$ is the density of states at Fermi energy, which we assume to be small but 
finite. 

Let us now specialize to the situation of interest to us, that describes Na$_3$Bi. 
Substituting Eq.~\eqref{eq:9} into Eq.~\eqref{eq:11}, we obtain 
\beq
\label{eq:13}
\bj = \frac{e^2}{\pi^2} \mu_5\, \bB + \tilde \sigma_{xy} (\hat z \times \tilde \bE)
\eeq
where $\tilde \sigma_{xy} = e^2 b /\pi^2$, and $\tilde \bE = - \bnabla \tilde A_0$. 
Analogously, 
\beq
\label{eq:14}
\tilde \bj = \frac{e^2}{\pi^2} \tilde \mu_5 \bB + \tilde \sigma_{xy} (\hat z \times \bE).
\eeq
The physical meaning of Eqs.~\eqref{eq:13} and \eqref{eq:14} is straightforward to understand. 
The first term on the right hand side in Eq.~\eqref{eq:13} describes the chiral magnetic effect~\cite{Kharzeev08,Son12}, i.e. 
a contribution to the charge current, proportional to the applied magnetic field and the chiral chemical potential. 
Analogously, the first term in Eq.~\eqref{eq:14} describes the $\mathbb Z_2$ magnetic effect, which generates a contribution to the 
spin current, proportional to the $\mathbb Z_2$ chemical potential and the external magnetic field.   
The second term in Eq.~\eqref{eq:14} describes the spin Hall effect, i.e. the generation of a spin current, transverse
to the applied external electric field, and $\tilde \sigma_{xy}$ is the corresponding spin Hall conductivity. 
Note that, in exact analogy to Weyl semimetals~\cite{Ran11,Burkov11-1}, this spin Hall conductivity may be associated 
with the Fermi arc edge states: each value of the momentum $k_z$ between the Dirac points contributes 
$e^2/\pi$ to the total spin Hall conductivity. 
Finally, the second term in Eq.~\eqref{eq:13} describes the inverse spin Hall effect, i.e. the generation of a charge current by 
a gradient of the spin density. 

Similarly, the chiral and the $\mathbb Z_2$ charge conservation laws take the following form
\beqa
\label{eq:15}
&&\frac{\partial n_5}{\partial t} + \bnabla \cdot \bj_5 = \frac{e^2}{2 \pi^2} \bE \cdot \bB - \frac{n_5}{\tau_5}, \nonumber \\
&&\frac{\partial \tilde n_5}{\partial t} + \bnabla \cdot \tilde \bj_5 = \frac{e^2}{2 \pi^2} \tilde \bE \cdot \bB 
- \frac{\tilde n_5}{\tau_5}, 
\eeqa
where we have introduced a finite relaxation time $\tau_5$ to account for the fact that the chiral $n_5$ and the 
$\mathbb Z_2$ $\tilde n_5$ charges 
are not in reality strictly conserved, when terms, nonlinear in momentum, are included in the Hamiltonian~\cite{Burkov_lmr_prb}. 
We have taken the relaxation times for both charges to be the same for simplicity. 
Note that the coefficient of $\bE \cdot \bB$ on the right hand side of the first of Eq.~\eqref{eq:15} is twice as large as it would be 
for a single Dirac point (or single pair of Weyl points). This is because we have two Dirac points (two pairs of Weyl points) 
and their chiral anomalies add up, which is not obvious in advance, the anomalies may cancel instead. 
The reason they add in our case is that there exists a conserved quantity (spin) that distinguishes the two pairs of Weyl fermions
and, as a result, the chiral chemical potential couples symmetrically to them.

To account for regular transport processes, not related to the anomalies, we also need to add ordinary Drude 
conductivity terms to the right hand side of Eqs.~\eqref{eq:13} and \eqref{eq:14}, which we do by hand~\cite{Son09}.
Then we finally obtain
\beqa
\label{eq:16}
\bj&=&\sigma_0 \bE+ \frac{e^2}{\pi^2} \mu_5\, \bB + \tilde \sigma_{xy} (\hat z \times \tilde \bE), \nonumber \\
\tilde \bj&=&\sigma_0 \tilde \bE + \frac{e^2}{\pi^2} \tilde \mu_5 \bB + \tilde \sigma_{xy} (\hat z \times \bE).
\eeqa
where $\sigma_0$ is the Drude conductivity. We take $\sigma_0$ to be the same for both charge and spin, which is true in our 
approximation of an exactly conserved spin. 

Eqs.~\eqref{eq:15} and \eqref{eq:16} must be solved simultaneously to obtain the magnetoresistance. 
Suppose a uniform DC charge current is injected into the sample along the $x$-direction, while voltage leads are attached 
to the edges, perpendicular to the $y$-direction. 
In this case we have from Eq.~\eqref{eq:15}
\beqa
\label{eq:17}
\mu_5 = \frac{n_5}{g(\epsilon_F)} = \frac{e^2 \tau_5}{2 \pi^2 g(\epsilon_F)} \bE \cdot \bB \nonumber \\
\tilde \mu_5 = \frac{\tilde n_5}{g(\epsilon_F)} = \frac{e^2 \tau_5}{2 \pi^2 g(\epsilon_F)} \tilde \bE \cdot \bB. 
\eeqa
Substituting these into Eq.~\eqref{eq:16}, we obtain
\beqa
\label{eq:18}
\bj&=&\sigma_0 \bE+ \chi (\bE \cdot \bB) \bB + \tilde \sigma_{xy} (\hat z \times \tilde \bE), \nonumber \\
\tilde \bj&=&\sigma_0 \tilde \bE + \chi (\tilde \bE \cdot \bB) \bB + \tilde \sigma_{xy} (\hat z \times \bE), 
\eeqa
where  $\chi = e^4 \tau_5 /2 \pi^4 g(\epsilon_F)$. 
We now solve the equations
\beq
\label{eq:19}
j_y = j_z = \tilde j_x = \tilde j_y = \tilde j_z = 0, 
\eeq
for $E_{y,z}, \tilde E_{x,y,z}$ and substitute the result into the equation for $j_x$ to obtain the diagonal resistivity
$\rho_{xx}$. The equation $\tilde j_x = 0$ follows from the assumption that we have spin-unpolarized leads. 
We obtain
\beq
\label{eq:20}
\rho_{xx}^{-1}(\bB) = \sigma_0 + \frac{\chi B_x^2 + \tilde \sigma_{xy}^2 \left(1 + \chi B_z^2/\sigma_0\right)/\sigma_0}
{1 + \chi (B_y^2 + B_z^2)/\sigma_0}. 
\eeq
Eq.~\eqref{eq:20} is our main result. 

\begin{figure}[t]
  \includegraphics[width=8cm]{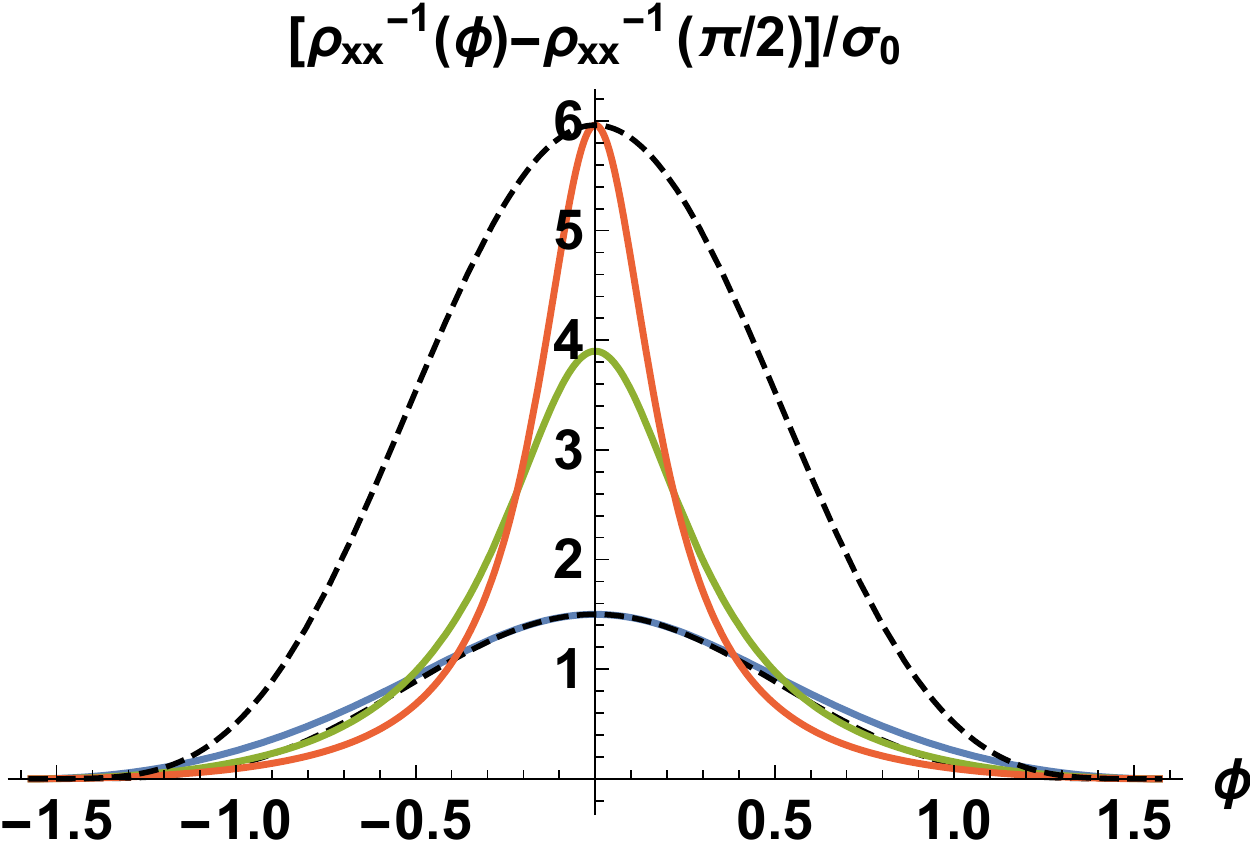}
  \caption{(Color online) Plots of $\rho^{-1}_{xx}(\phi) - \rho^{-1}_{xx}(\pi/2)$ for different values of the magnetic 
  field, assuming $\tilde \sigma_{xy} = \sigma_0$ and for the field rotated in the $xy$-plane. 
  Dashed lines represent $\cos^4 \phi$ dependence, normalized to the same maximum value. 
  The fit to $\cos^4 \phi$ is good for the smallest magnitude of the magnetic field (solid blue line), but becomes very poor for the largest (solid red line).} 
  \label{fig:2}
\end{figure}

Several features of Eq.~\eqref{eq:20} are noteworthy. 
Setting $\bB = 0$ we obtain
\beq
\label{eq:21}
\rho_{xx}^{-1}(0) = \sigma_0 + \frac{\tilde \sigma^2_{xy}}{\sigma_0}. 
\eeq
The second term in Eq.~\eqref{eq:21} represents a reduction of the diagonal resistivity due to the spin Hall effect, which in 
turn is associated with the Fermi arc surface states.
At a nonzero magnetic field, the dependence of $\rho_{xx}^{-1}(\bB)$ on the angle between the magnetic field and the current
demonstrates the narrowing effect, observed in Ref.~\cite{Ong_anomaly}, see Fig.~\ref{fig:2}. 
The origin of this effect is the magnetic field dependence of the denominator in Eq.~\eqref{eq:20}. 
Physically, this follows from the magnetic field dependence of the chiral chemical potential, Eq.~\eqref{eq:17}. 
Another consequence of the $\mathbb Z_2$ anomaly (and the quantum spin Hall effect as one of its manifestations), is the anisotropy between the angular dependences of the magnetoresistance when the magnetic field is rotated in the $xy$- and the $xz$-planes. 
Let $\phi$ be the angle between the magnetic field and the $x$ axis when the field is rotated in the $xy$-plane while 
$\theta$ be the same angle when the field is rotated in the $xz$-plane. Then we obtain
\beq
\label{eq:22}
\rho_{xx}^{-1}(\theta = \phi) - \rho_{xx}^{-1}(\phi) = \frac{\tilde \sigma_{xy}^2}{\sigma_0} \frac{\chi B^2 \sin^2 \phi/ \sigma_0}
{1 + \chi B^2 \sin^2 \phi/ \sigma_0}, 
\eeq
where $\rho_{xx}^{-1}(\theta)$ refers to the inverse resistivity for the field rotated in the $xz$-plane, while $\rho_{xx}^{-1}(\phi)$ refers to the inverse resistivity for the field rotated in the $xy$-plane. 
This anisotropy exists only when $\tilde \sigma_{xy}$ is not zero and is a direct consequence of the spin Hall effect and thus the 
$\mathbb Z_2$ anomaly. 

In conclusion, we have demonstrated that in Dirac semimetals with two Dirac nodes, separated in momentum space along a 
rotation axis and characterized by a nontrivial $\mathbb Z_2$ topological charge, there exist the corresponding $\mathbb Z_2$ anomaly. 
This refers to anomalous nonconservation of the $\mathbb Z_2$ topological charge in the presence of external electromagnetic 
fields and gradients of the (nearly) conserved spin density and is closely analogous to the chiral anomaly, which is also present. 
We have shown that the interplay of the $\mathbb Z_2$ and the chiral anomalies leads to observable effects in magnetotransport. 
We have also provided a possible explanation for the magnetic field dependent narrowing of the dependence of the positive magnetoconductivity on the angle between the current and the applied magnetic field, observed in a recent 
experiment~\cite{Ong_anomaly}.  
\begin{acknowledgments}
Financial support was provided by NSERC of Canada. 
\end{acknowledgments}
\bibliography{references}
\end{document}